\documentclass{pos}

\usepackage{bm}
\usepackage{amsmath}
\usepackage{rsfs}
\newcommand{\preprint}{
  \begin{picture}(0,0)
    \put(140,120){{\rm\normalsize CHIBA-EP-174}}
  \end{picture}}

\title{Magnetic monopole loops supported by a meron pair as the quark confiner\preprint}

\ShortTitle{Magnetic monopole loops by a meron pair}

\author{Kei-Ichi Kondo\thanks{This
work is financially supported by Grant-in-Aid for
Scientific Research (C) 18540251 from Japan Society for
the Promotion of Science (JSPS).}\\
        Department of Physics, Graduate School of Science,  Chiba University, Chiba 263-8522, Japan
\\
        E-mail: \email{kondok@faculty.chiba-u.jp}}

\abstract{
We give a definition of gauge-invariant magnetic monopoles in Yang-Mills theory without using the Abelian projection due to  't Hooft.  They automatically appear  from the Wilson loop operator. This is shown by rewriting the Wilson loop operator using a non-Abelian Stokes theorem.  
The magnetic monopole defined in this way is a topological object of co-dimension 3, i.e., a loop in four-dimensions. We show that such magnetic loops indeed exist in four-dimensional  Yang-Mills theory.
In fact, we give an analytical solution representing circular magnetic monopole loops joining a pair of merons in the four-dimensional Euclidean SU(2) Yang-Mills theory. 
This is achieved by solving the differential equation for the adjoint color (magnetic monopole) field  in the two--meron background field  within the recently developed reformulation of the Yang-Mills theory. 
Our analytical solution corresponds to the numerical solution found by Montero and Negele on a lattice. 
This result strongly suggests that a meron pair is the most relevant quark confiner in the original Yang-Mills theory, as Callan, Dashen and Gross suggested long ago. 
}

\FullConference{8th Conference Quark Confinement and the Hadron Spectrum \\
		 September 1-6 2008\\
		 Mainz, Germany}

\begin{document}

\section{Wilson loop and magnetic monopole}

For a closed loop $C$, the Wilson loop operator for SU(2) Yang-Mills connection is defined by
\begin{eqnarray}
  W_C[\mathbf{A}]  
:=& {\rm tr} \left[ \mathscr{P} \exp \left\{ ig \oint_{C} dx^\mu \mathbf{A}_\mu(x) \right\} \right]/{\rm tr}({\bf 1})  
 ,
\quad 
  \mathbf{A}_\mu(x)=\mathbf{A}_\mu^A(x) \sigma^A/2  .
\end{eqnarray}
The path-ordering $\mathscr{P}$ is removed by using the Diakonov-Petrov version \cite{DP89} of a non-Abelian Stokes theorem  for the  Wilson loop operator: in the $J$ representation of SU(2) ($J=1/2, 1, 3/2, 2, \cdots$)  
\begin{center}
\begin{picture}(-320,-2950)
\put(0,-20){\includegraphics[height=2.5cm]{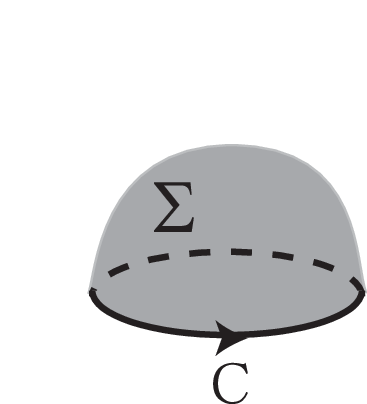}}%
\end{picture}
\end{center}
\vskip -1.5cm
\begin{eqnarray}
  W_C[\mathbf{A}]  
&:=&  \int  d\mu_\Sigma(U) \exp \left\{ i  Jg \int_{\Sigma: \partial \Sigma=C} dS^{\mu\nu} f_{\mu\nu} \right\} , 
 \ \text{no path-ordering}
  \nonumber\\
  f_{\mu\nu}(x) &:=& \partial_\mu [\mathbf{A}^A_\nu(x)\bm{n}^A(x)  ] -  \partial_\nu [\mathbf{A}^A_\mu(x)\bm{n}^A(x)  ]   
  - g^{-1} \epsilon^{ABC} \bm{n}^A(x)  \partial_\mu \bm{n}^B(x)  \partial_\nu \bm{n}^C(x) 
 ,
\nonumber\\
 n^A(x) \sigma^A  &:=& U^\dagger(x)  \sigma^3 U(x) , \quad U(x) \in SU(2) \quad (A,B,C \in \{ 1,2,3 \}) ,
 \label{n}
\end{eqnarray}
where $d\mu_\Sigma(U)$ is the product measure of an invariant measure on SU(2)/U(1) over $\Sigma$: 
\begin{eqnarray}
 d\mu_\Sigma(U) :=\prod_{x \in \Sigma}d\mu(U(x)) ,
  \quad
  d\mu(U(x)) = {2J+1 \over 4\pi} \delta(\bm{n}^A(x)   \bm{n}^A(x)-1) d^3 \bm{n}(x)
 ,  
\end{eqnarray}
where we have introduced a unit vector field $\bm{n}(x)$. 

The geometric and topological meaning of the Wilson loop operator was given in \cite{Kondo08}:
\begin{eqnarray}
  W_C[\mathscr{A}]  =& \int  d\mu_\Sigma(U)  \exp \left\{  iJg (\Xi_{\Sigma}, k)  + iJg (N_\Sigma,j)  \right\} , 
  \quad 
  C = \partial \Sigma
\\
  &  k:= \delta {}^*f ={}^*df, \quad
  \Xi_\Sigma := \delta {}^*\Theta_\Sigma \triangle^{-1}  \leftarrow \quad \text{(D-3)-forms}
\\
  & j:= \delta f, \quad
  N_\Sigma := \delta \Theta_\Sigma \triangle^{-1}   \leftarrow \quad \text{1-forms (D-indep.)}
\\
  & \Theta^{\mu\nu}_\Sigma(x) = \int_{\Sigma} d^2S^{\mu\nu}(x(\sigma)) \delta^D(x-x(\sigma))
   ,
\end{eqnarray}
where 
$k$ and $j$ are gauge invariant and conserved currents, $\delta k=0=\delta j$.
Thus,  {\bf  
we do not need to use the Abelian projection proposed by 't Hooft \cite{tHooft81} to define   magnetic monopoles in Yang-Mills theory!
}
 {\bf  
The Wilson loop operator knows the (gauge-invariant) magnetic monopole!
}
Then the magnetic monopole is a topological object of co-dimension 3. 
In $D$ dimensions, 
\\ 
D=3:  0-dimensional point defect $\rightarrow$ magnetic monopole of Wu-Yang type
\\
D=4: 1-dimensional line defect  $\rightarrow$  magnetic monopole loop (closed loop)

For $D=3$, 
\begin{equation}
k(x)=\frac12 \epsilon^{jk\ell}\partial_\ell f_{jk}(x)=\rho_m(x)
\end{equation}
 denotes the magnetic charge density at $x$, and 
\begin{equation}
\Xi_{\Sigma}(x)=\Omega_{\Sigma}(x)/(4\pi)
\end{equation}
 agrees with the (normalized) solid angle  at the point $x$ subtended by the surface $\Sigma$ bounding the Wilson loop $C$. 
 Then the magnetic part $W_{\mathscr{A}}^m$ is written as 
\begin{picture}(-500,-2000)
\put(90,30)
{\includegraphics[height=2.5cm]{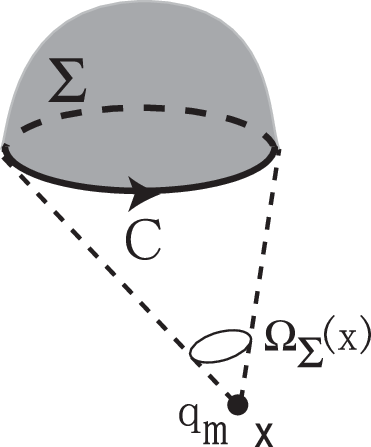}}
\end{picture}
\begin{equation}
 W_{\mathscr{A}}^m := \exp \left\{  iJg (\Xi_{\Sigma}, k)    \right\}
= \exp \left\{ iJg \int d^3x \rho_m(x)  \frac{\Omega_{\Sigma}(x)}{4\pi}  \right\} .
\end{equation}
The magnetic charge $q_m$ obeys the Dirac-like quantization  condition:
\begin{equation}
 q_m :=  \int d^3x \rho_m(x) =4\pi g^{-1} n \quad (n \in \mathbb{Z}) .
\end{equation}
The proof follows from a fact that the non-Abelian Stokes theorem does not depend on the surface $\Sigma$ chosen for spanning the surface bounded by the loop $C$.  See \cite{Kondo08}.

 For an  ensemble of point-like magnetic charges:
$
 k(x) = \sum_{a=1}^{n} q_m^a \delta^{(3)}(x-z_a) 
$, we have
\begin{equation}
W_{\mathscr{A}}^m  
= \exp \left\{  iJ\frac{g}{4\pi} \sum_{a=1}^{n} q_m^a \Omega_\Sigma(z_a)   \right\}
= \exp \left\{  iJ  \sum_{a=1}^{n} n_a \Omega_\Sigma(z_a)   \right\}
 , \quad n_a \in \mathbb{Z} .
\end{equation}
The magnetic monopoles in the neighborhood of the Wilson surface $\Sigma$ ($\Omega_\Sigma(z_a)= \pm 2\pi$) contribute to the Wilson loop 
\begin{equation}
W_{\mathscr{A}}^m  
= \prod_{a=1}^{n} \exp ( \pm i2 \pi J n_a )
= \begin{cases}
 \prod_{a=1}^{n} (-1)^{n_a} &(J=1/2,3/2, \cdots) \cr
 = 1 &(J=1,2,\cdots) 
 \end{cases} .
\end{equation}
This enables us to explain the $N$-ality dependence of the asymptotic string tension.  See, \cite{Kondo08b}.


For $D=4$, $\Omega_{\Sigma}^\mu(x)$ is the $D=4$  solid angle and the magnetic part reads 
\begin{equation}
W_{\mathscr{A}}^m  
= \exp \left\{ iJg  \int d^4x \Omega_{\Sigma}^\mu(x) k^\mu(x)   \right\}
.
\end{equation}
Suppose the existence of an ensemble of  magnetic monopole loops $C_a^\prime$ in $D=4$ Euclidean space,
$
 k^\mu(x)  =  \sum_{a=1}^{n} q_m^a \oint_{C_a^\prime} dy^\mu_a \delta^{(4)}(x-x_a) 
 , \quad
 q_m^a = 4\pi g^{-1} n_a  
$.
Then the Wilson loop operator reads
\begin{equation}
W_{\mathscr{A}}^m  
 =  \exp \left\{  iJg \sum_{a=1}^{n} q_m^a L(\Sigma, C_a^\prime)  \right\}
= \exp \left\{   4\pi J i \sum_{a=1}^{n} n_a L(\Sigma, C_a^\prime)   \right\}
 , \quad n_a \in \mathbb{Z} 
,
\end{equation}
where
$L(\Sigma, C^\prime)$ is the  {linking number} between the surface $\Sigma$  and  the curve $C^\prime$: 
\begin{center}
\begin{picture}(-360,-3800)
\put(-70,-20){\includegraphics[height=1.5cm]{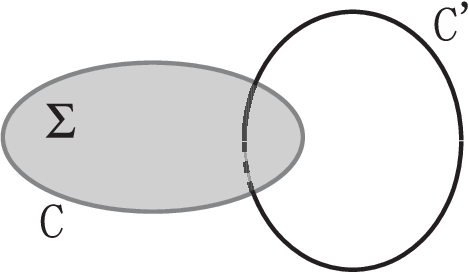}}%
\end{picture}
\end{center}
\vskip -0.9cm
\begin{equation}
  L(\Sigma, C^\prime)  
  := \oint_{C^\prime} dy^\mu(\tau) \Xi^\mu_{\Sigma}(y(\tau)) 
.
\end{equation} 
Here the curve $C^\prime$ is identified with the trajectory  $k$ of a magnetic monopole and the surface $\Sigma$ with the world sheet of a hadron (meson) string for a quark-antiquark pair.

The Wilson loop operator is a probe of the gauge-invariant magnetic monopole defined in our formulation.
Thus, calculating the Wilson loop average reduces to the summation over the magnetic monopole charge (D=3) or current (D=4) with a geometric factor, the solid angle (D=3) or linking number (D=4).

\section{Main results (Magnetic loops indeed exist in YM$_4$)}

We can show that {\bf the gauge-invariant magnetic loop (assumed in the above) indeed exists in SU(2) Yang-Mills theory in $D=4$ Euclidean space}: we give a first* (exact) 
analytical solution representing {\bf circular magnetic monopole loops joining two merons} \cite{KFSS08}.%
\footnote{There is an exception: 
Bruckmann \& Hansen,
hep-th/0305012,
Ann.Phys.{\bf 308}, 201 (2003).  However,  it has $Q_P=\infty$
}
\begin{center}
\begin{picture}(-350,-3000)
\put(-40,-30){\includegraphics[height=2cm]{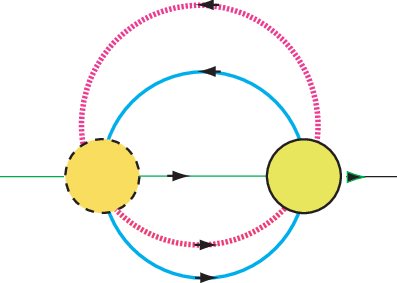}}%
\end{picture}
\end{center}

\vskip 0.7cm
Our method reproduces also the previous results based on MAG (MCG) and LAG:
\\
\noindent
(i) The magnetic straight line can be obtained in the one-instanton or one-meron background. \cite{CG95,RT01}

\noindent
(ii) The magnetic closed loop can NOT be obtained in the one-instanton background. \cite{BOT97,BHVW01}
\\

\section{Reformulating Yang-Mills theory in terms of new variables}

\noindent
SU(2) Yang-Mills theory $\quad\quad\quad\quad$
$\quad\quad\quad$ A reformulated Yang-Mills theory 
\\
written in terms of 
$\quad\quad\quad\quad$
$\Longleftrightarrow$ 
$\quad\quad\quad$
 written in terms of new variables:
\\
$\mathbf{A}_\mu^A(x)$  $(A=1,2,3)$
$\quad$ 
change of variables
$\quad\quad$
$\bm{n}^A(x), c_\mu(x), \mathbf{X}_\mu^A(x)$
$(A=1,2,3)$

We introduce a ``color field'' ${\bf n}(x)$ of unit length with three components
\begin{eqnarray}
{\bf n}(x) = (n_1(x),n_2(x),n_3(x)) , 
\quad
 {\bf n}(x) \cdot {\bf n}(x) = n_A(x) n_A(x) = 1 
\end{eqnarray}
The color field ${\bf n}(x)$ is identified with ${\bf n}(x)$ in (\ref{n}).  New variables $\bm{n}^A(x), c_\mu(x), \mathbf{X}_\mu^A(x)$  should be given as functionals of the original $\mathbf{A}_\mu^A(x)$. 
The off-shell  {Cho-Faddeev-Niemi-Shabanov decomposition} \cite{CFNS} is reinterpreted as  { change of variables from $\mathbf{A}_\mu^A(x)$ to $\bm{n}^A(x), c_\mu(x), \mathbf{X}_\mu^A(x)$} via the reduction of an enlarged gauge symmetry.
See \cite{KMS06,KSM08}.
Expected role of the color field:
1) The color field $\bm{n}(x)$ plays the role of recovering color symmetry which will be lost 
in the conventional approach, e.g., in the MA gauge. 
2) The color field $\bm{n}(x)$  carries topological defects responsible for non-perturbative phenomena, e.g., quark confinement. 

\section{Bridge between $\mathbf{A}_\mu(x)$  and $\mathbf{n}(x)$}

For a given Yang-Mills field $\mathbf{A}_\mu(x)$,  the {  color field} $\mathbf{n}(x)$ is obtained by solving the   {  reduction differential equation (RDE)}: \cite{KSM08}
\begin{equation}
 \mathbf{n}(x) \times D_\mu[\mathbf{A}]D_\mu[\mathbf{A}] \mathbf{n}(x) = \mathbf{0} 
 .
\end{equation}

 For a given SU(2) Yang-Mills field $\mathbf{A}_\mu(x)=\mathbf{A}^A_\mu(x)\frac{\sigma_A}{2}$,  look for  unit vector fields $\mathbf{n}(x)$ such that   $-D_\mu[\mathbf{A}] D_\mu[\mathbf{A}]\mathbf{n}(x)$ is proportional to $\mathbf{n}(x)$:  
 an eigenvalue-like form: 
\begin{align}
 -D_\mu[\mathbf{A}]D_\mu[\mathbf{A}] \mathbf{n}(x) = \lambda(x) \mathbf{n}(x) 
 \quad (\lambda(x) \ge 0) .
\end{align}
The solution is not unique. 
We choose the solution giving the smallest value of the reduction functional $F_{\rm rc}$ which agrees with the  integral of the scalar function $\lambda(x)$ over $\mathbb{R}^D$F
\begin{align}
 F_{\rm rc} =&  \int d^Dx \frac12 (D_\mu[\mathbf{A}] \mathbf{n}(x)) \cdot (D_\mu[\mathbf{A}] \mathbf{n}(x)) 
 =   \int d^Dx \frac12  \mathbf{n}(x) \cdot (-D_\mu[\mathbf{A}] D_\mu[\mathbf{A}] \mathbf{n}(x)) 
  \nonumber\\
\Longrightarrow F_{\rm rc}^* =&  \int d^Dx \frac12  \mathbf{n}(x) \cdot \lambda(x) \mathbf{n}(x) 
 =  \int d^Dx  \frac12  \lambda(x)
 .
\end{align}

\section{Conclusion and discussion}

For given one-instanton and two-meron background   $\mathbf{A}_\mu(x)$, we have solved the RDE for the color field ${\bf n}(x)$ \cite{KSM08}. 
In the four-dimensional Euclidean SU(2) Yang-Mills theory,
we have given a first 
{ 
 analytical solution representing circular magnetic monopole loops $k_\mu$ which go through a pair of merons 
  (with a unit topological charge)} with non-trivial linking with the Wilson surface $\Sigma$.

\begin{center}
\begin{picture}(-350,-3000)
\put(-150,-30){
\includegraphics[width=5cm]{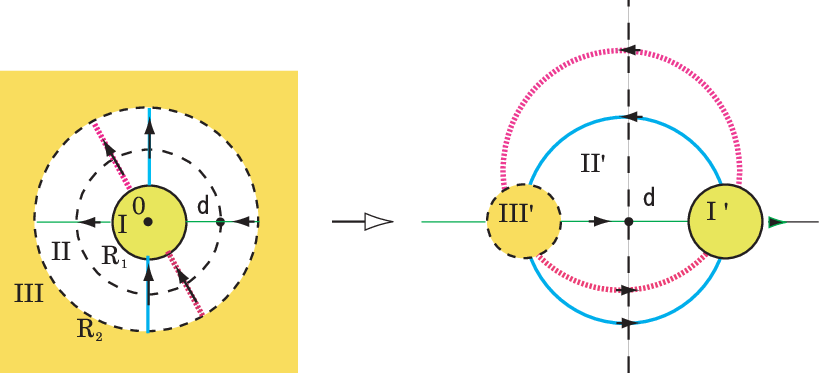}
}%
\end{picture}
\end{center}
\noindent
This is achieved by solving the reduction differential equation for the adjoint color (magnetic monopole) field  in the two--meron background field using the  recently developed reformulation of the Yang-Mills theory  \cite{KMS06,KSM08} and a non-Abelian Stokes theorem \cite{Kondo08}. 

Our analytical solution corresponds to a numerical solution found  on a lattice by Montero and  Negele \cite{MN02}.

\begin{center}
\begin{picture}(-350,-3000)
\put(-300,-60){
\includegraphics[width=3.5cm]{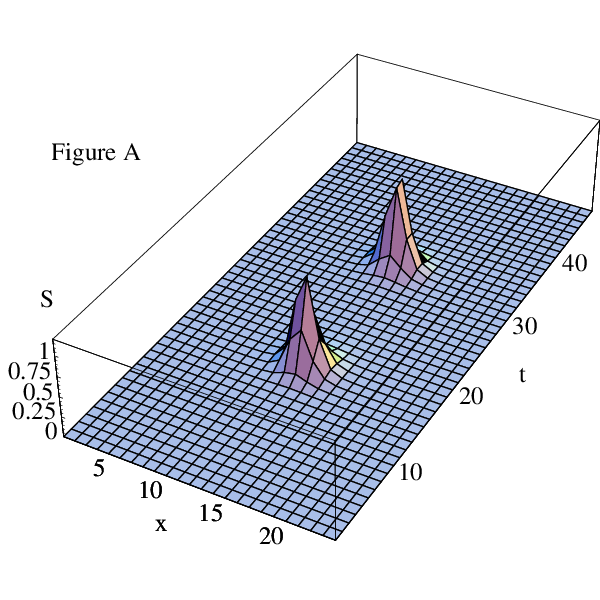}
\includegraphics[width=3.5cm]{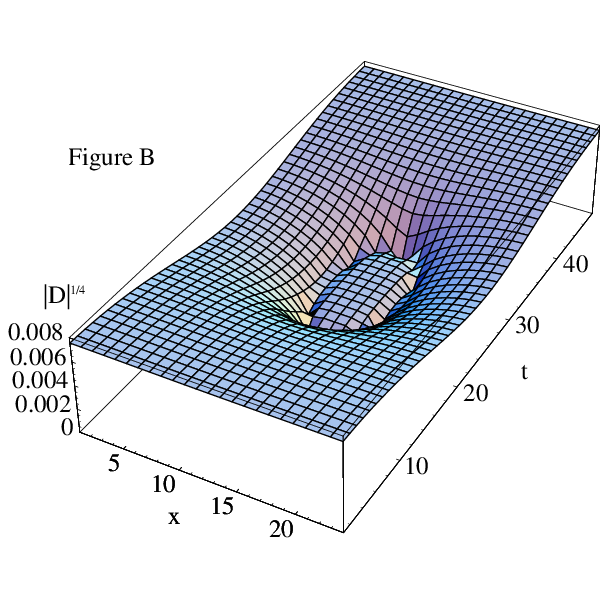}
}%
\end{picture}
\end{center}

\vskip 1.5cm
We have not yet obtained the analytic solution representing magnetic loops connecting 2-instantons, which were found in the numerical way by Reinhardt \& Tok \cite{RT01}.
\begin{center}
\begin{picture}(-350,-3000)
\put(-300,-80){
\includegraphics[width=2.5cm]{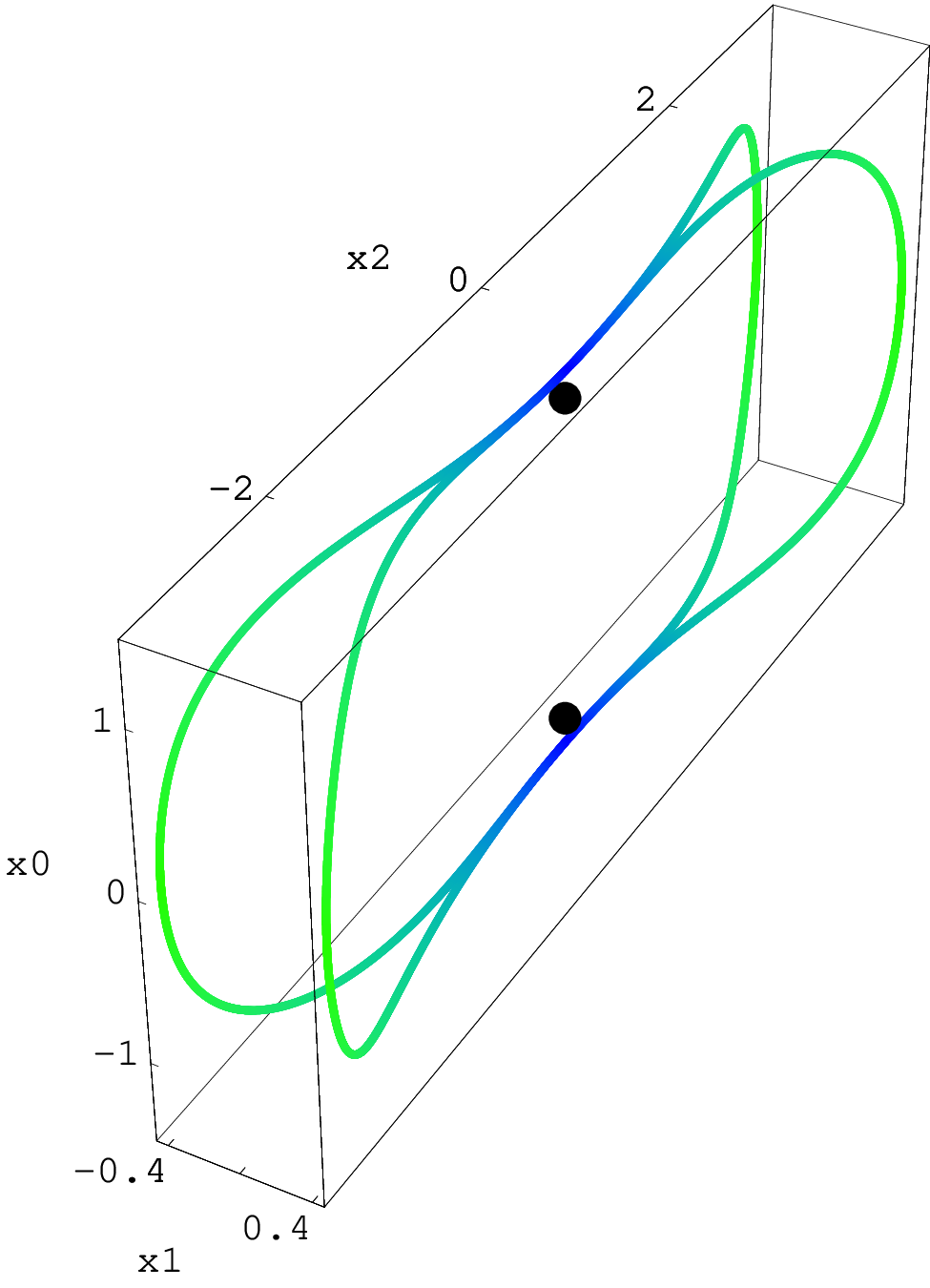}
}%
\end{picture}
\end{center}

\vskip 2.5cm
Thus we are lead to a conjecture: 
A meron pair is the most relevant quark confiner in the original Yang-Mills theory, as Callan, Dashen and Gross suggested long ago \cite{CDG78}.  This means a duality relation:
\begin{center}
dual Yang-Mills: magnetic monopole loops
$\Longleftrightarrow$ original Yang-Mills: merons 
\end{center}

\end{document}